\documentclass[pdflatex,sn-mathphys-num]{sn-jnl}

\usepackage{graphicx}%
\usepackage{multirow}%
\usepackage{amsmath,amssymb,amsfonts}%
\usepackage{amsthm}%
\usepackage{mathrsfs}%
\usepackage[title]{appendix}%
\usepackage{xcolor}%
\usepackage{textcomp}%
\usepackage{manyfoot}%
\usepackage{booktabs}%
\usepackage{algorithm}%
\usepackage{algorithmicx}%
\usepackage{algpseudocode}%
\usepackage{listings}%

\theoremstyle{thmstyleone}%
\theoremstyle{thmstyletwo}%

\theoremstyle{thmstylethree}%

\begin{document}

\title[Mass spectrum of strange hidden-charm pentaquarks]{$P_{c\bar cs}(4459)^{0}$, $P_{c\bar c s}(4338)^0$ and mass spectrum of strange hidden-charm pentaquarks}


\author[1]{\fnm{Zhe-Hao} \sur{Cao}}

\author[1]{\fnm{Zhi-Yuan} \sur{Chen}}

\author[2]{\fnm{You-You} \sur{Lin}}

\author[1]{\fnm{Ji-Ying} \sur{Wang}}

\author*[1]{\fnm{Ailin} \sur{Zhang}}\email{zhangal@shu.edu.cn}

\affil[1]{\orgdiv{Department of Physics}, \orgname{Shanghai University}, \orgaddress{\city{Shanghai}, \postcode{200444}, \country{China}}}
\affil[2]{\orgdiv{School of Electronic Information}, \orgname{Huzhou College}, \orgaddress{\city{Huzhou}, \postcode{313000}, \country{China}}}

\abstract{Strange hidden-charm pentaquark states have been systematically investigated within a diquark-triquark model. Through a Gaussian expansion method, masses of some diquarks, triquarks and strange hidden-charmed pentaquark states from S-wave to P-wave excitations have been calculated with the non-relativistic Semay and Silvestre-Brac potentials in terms of the same parameters employed for tetraquark states. Masses of pentaquark states in S-wave excitations are found between $4200$ MeV and $4590$ MeV, while masses of all P-wave excitations are found above $4600$ MeV. Mass splittings between the S-wave and P-wave pentaquark states are about $350-570$ MeV. In comparison to the experimental data, $P_{c\bar cs}(4459)^{0}$ observed by LHCb in decay channel $\Xi_{b}^{-}\rightarrow J/\psi \Lambda K^-$ is assumed as the $|1; 0, 1/2; 3/2, 0\rangle_{3/2}$ $[sq][\bar{c}cq]$ pentaquark state with $J^P={3\over 2}^-$, while $P_{c\bar c s}(4338)^0$ observed in the decay channel $B^{-}\rightarrow J/\psi \Lambda \bar{p}$ is very possibly the $|0; 1, 1/2; 1/2, 0\rangle_{1/2}$ $[cq][\bar{c}sq]$ pentaquark state with $J^P={1\over 2}^-$. We predict a lowest strange hidden-charm pentaquark state with $J^P={1\over 2}^-$ around $4200$ MeV.}

\maketitle

\section{Introduction}

A study of hadron spectroscopy has been a crucial approach to understand strong interactions, particularly to explore the non-perturbative properties of Quantum Chromodynamics (QCD) at low energy and the mechanism of quark confinement. In the naive quark model proposed by Gell-Mann and Zweig ~\cite{Gell-Mann:1964ewy,Zweig:1964jf}, conventional baryons are composed of three quarks or three antiquarks, while conventional mesons consist of a quark and an antiquark. Of course, theory also allows for the existence of hadronic states with more quark constituents. In 1977, Jaffe investigated $qq\bar{q}\bar{q}$ multiquark states using the MIT bag model~\cite{PhysRevD.15.267}; in 1979, Strottman studied the $qqqq\bar{q}$ configuration within the same model~\cite{Strottman:1979qu}. Lipkin first introduced the term "pentaquark" and predicted an anticharmed strange baryonic state with the composition $\bar{c}suud$~\cite{Lipkin1987NewPF} in 1987. A pentaquark consists of four quarks and an antiquark according to the $\mathbf{SU(3)}$ colour confinement. More references on the study of pentaquark states could be found in some reviews and references therein~\cite{pr639.1,pr668.1,ppnp93.143,rmp90.015003,rmp90.015004,pr873.1}. 

A light pentaquark state $\Theta^+$ with mass $\sim 1540$ MeV was announced by LEPS in 2003~\cite{prl91.012002}, but this state has not been confirmed by subsequent experiments. A major breakthrough in search for pentaquarks was made in 2015. The LHCb collaboration observed hidden-charm pentaquark states consistent with the $c\bar{c}uud$ quark configuration in the decay $\Lambda_{b}^{0}\rightarrow J/\psi p K^-$~\cite{PhysRevLett.115.072001, PhysRevLett.117.082002}. Initial analysis revealed a broader structure, $P_{c}(4380)^+$, and a narrower resonance, $P_{c}(4450)^+$. In 2019, with a larger data sample, LHCb observed fine structures in the same decay channel: the peak previously seen at 4450 MeV was resolved into two narrowly separated resonances, $P_{c}(4440)^{+}$ and $P_{c}(4457)^{+}$. Besides, a new narrow resonance, $P_{c}(4312)^{+}$, was also discovered~\cite{PhysRevLett.122.222001}.

Based on a $SU(3)$ flavor symmetry, the existence of hidden-charm pentaquark states containing a strange quark was theoretically anticipated \cite{PhysRevLett.105.232001, PhysRevC.84.015202}, and searches in the $J/\psi\Lambda$ final state of $\Xi _{b}^{-}\rightarrow J/\psi \Lambda K^- $ decays were suggested~\cite{PhysRevC.93.065203}. Using the full dataset from Run 1 and Run 2, the LHCb experiment reconstructed approximately 1,750 signal events of $\Xi _{b}^{-}\rightarrow J/\psi \Lambda K^- $~\cite{lhcb2021evidence}. Amplitude analysis demonstrated that inclusion of the contributions from strange hidden-charm pentaquark states had significantly improved the fit to data. This experiment marked the first evidence for a strange hidden-charm pentaquark state, named $P_{c\bar cs}(4459)^{0}$, with a mass of $4458.8 \pm 2.9_{-1.1}^{+4.7}$ MeV and a decay width of $17.3\pm 6.5_{-5.7}^{+8.0}$ MeV. This state has a mass about 19 MeV below the $\Xi_c \bar{D}^*$ threshold and a narrow width.

Subsequently, the LHCb experiment reconstructed the $B^{-}\rightarrow J/ \psi \Lambda \bar{p}$ decay channel through the same dataset. Once again, amplitude analysis confirmed that incorporating contributions from strange hidden-charm pentaquark states was essential for an accurate description of the data. In fact, the experiment led to a first observation of a strange hidden-charm pentaquark state $P_{c\bar c s}(4338)^0$ with minimal quark content $c\bar{c}uds$. The statistical significance for this state reached 15 standard deviations. Its mass and width were measured to be $4338.2 \pm 0.7 \pm 0.4$ MeV and $7.0\pm 1.2 \pm 1.3$ MeV, respectively. Its spin quantum number was determined with $J=1/2$, while its parity has not yet been definitively established though a negative parity is favored by the experimental data~\cite{PhysRevLett.131.031901}.

For an interpretation of the light strange pentaquark, two popular structures have been proposed: a pentaquark state may consist of two scalar diquarks in a relative P-wave and a strange antiquark~\cite{prl91.232003}, it may consist of a triquark and a diquark in a relative P-wave~\cite{plb575.249}. So far, the hidden-charmed pentaquark states have been extensively studied. Regarding their interior structure and dynamics, the hidden-charmed pentaquark state may consist of two diquarks and an antiquark~\cite{MAIANI2015289,Ghosh:2015xqp}, a triquark and a diquark~\cite{LEBED2015454,plb756.259,PhysRevD.104.114028}, or a $c\bar c$ cluster with a $qqs$ ($q=u,~d$) cluster (molecular pentaquark)~\cite{PhysRevLett.105.232001, PhysRevC.84.015202,PhysRevLett.115.122001,PhysRevLett.115.132002,Huang:2015uda,PhysRevD.92.094003,plb753.547,prd97.034006,prl122.242001}. The observed hidden-charmed pentaquark candidates were also explained as bound D-soliton states~\cite{prd92.051501}, anomalous triangle singularity~\cite{plb757.231}, kinematical effects~\cite{PhysRevD.92.071502} or compact states~\cite{prd96.014014}. No matter what it is, a pentaquark state as a system with four quarks and an antiquark has complicated interior structure and dynamics, which is not clear yet and deserves more investigation.

In a constituent diquark-antidiquark model, masses for different kinds of tetraquark states have been computed with the nonrelativistic potentials proposed by Semay and Silvestre-Brac~\cite{zpc61.271,fbs20.1}. The predicted masses are consistent with some observed tetraquark candidates very well~\cite{prd111.014015,prd112.034036,arxiv.2602.11744}. In this paper, we will concentrate on strange hidden-charm pentaquark states in a diquark-triquark configuration and compute their masses in terms of the Semay and Silvestre-Brac potential with only the same parameters employed in the study of tetraquarks. 

The paper is organized as follows, a simple introduction to the diquark-triquark potential model is given in Sec. II,  numerical calculation and some analyses are presented in Sec. III. The final section is devoted to a brief summary and discussion.

\section{Constituent diquark-triquark potential model}
In the diquark-triquark model proposed by Karliner and Lipkin~\cite{plb575.249}, the $ud$ diquark cluster is kept together with the $ud\bar s$ triquark cluster in a relative P-wave by the color electric force, and the color hyperfine interaction operates only within each cluster. The $ud$ diquark is in the $\bar 3_c$ of the color $SU(3)_c$ and in the $\bar 3_f$ of the flavor $SU(3)_f$, and has $I=0$, $S=0$. The triquark $ud\bar s$ consists of the diquark and antiquark coupled to an $SU(3)_c$ triplet $3_c$. It is in a $\bar 6$ of the flavor $SU(3)_f$ and has $I=0$, $S=1/2$.

In this paper, a pentaquark is assumed as a bound system with a diquark and a triquark, and the triquark consists of a diquark and antiquark without relative excitation. There are possible orbital excitations between the diquark and the triquark, the interaction between the diquark and the triquark is assumed as the same as that between the quark and the antiquark inside a normal meson. A diquark can be in a color antitriplet or color sextet state for the decomposition $3\otimes 3= \bar{3} \oplus 6$. A triquark may be in a $3$, $\bar{6}$, $3$ or $15$ color representation for the following decomposition $3 \otimes 3 \otimes \bar{3} = (\bar{3} \oplus 6) \otimes \bar{3} = 3 \oplus \bar{6} \oplus 3 \oplus 15$. 

Regarding the interaction between quarks, the binding of $q\bar q$ or $qq$ system depends on the quadratic Casimir invariants through a discriminant factor
\begin{align}
	I=\frac{1}{2}(C_{2}(R)-C_{2}(R_{1})-C_{2}(R_{2}))
\end{align}
where $C_{2}(R)$ is the quadratic Casimir invariant for the system's representation $R$, while $C_{2}(R_{1})$ and $C_{2}(R_{2})$ are the quadratic Casimir invariants for the component's representations $R_{i}$ \cite{PhysRevLett.113.112001,plb756.259}. When $I$ is negative, the particles attract each other. 

For representations $R=(1,\bar{3},6,8)$, the corresponding values are $I=\frac{1}{6}(-8,-4,+2,+1)$. Therefore, only the color singlet state exhibits attractive color forces for a $q\bar q$ system, while the color antitriplet state exhibits attractive color force for a $qq$ system. The color attraction in the $q\bar q$ system is twice strong as that in the $qq$ system. 

For the triquark $q_1q_2\bar q_3$, there are two cases with color attraction based on the decomposition $R=(3,\bar{6},3,15)$. In the first color triplet case, $\bar{q}_{3}$ attracts the $q_{1}q_{2}$ pair, while $q_{1}$ and $q_{2}$ repel each other. In the second color triplet case, $\bar{q}_{3}$ attracts the $q_{1}q_{2}$ pair, but $q_1$ and $q_2$ attract each other unlike the first case.
 
Inside a binding system, the relativistic energy of a quark is
\begin{equation}
    \begin{aligned}
    \label{kinetic term}
    E &=\sqrt{p^{2}+m^{2}} \\
    &=m(1+\frac{p^{2}}{2m^{2}}-\frac{p^{4}}{8m^{4}}+\cdots).
    \end{aligned}
\end{equation}
In a non-relativistic approximation
\begin{align}    
 E  &\approx m+\frac{p^{2}}{2m}
\end{align}
with the neglect of higher-order terms.

The Hamiltonian of the system is expressed as
$$H= \sum_{i}(m_{i}+\frac{p_{i}^{2}}{2m_{i}})+\sum_{i<j}V(r_{ij}).$$

Therefore, the corresponding Hamiltonian for pentaquark states can be written as
\begin{equation}
    \begin{aligned}
    \label{Hamiltonian}
    H&= m_{1}+m_{2}+\frac{p^{2}}{2\mu}+V(\textbf{r})
    \end{aligned}
\end{equation}
where $m_1$ and $m_2$ are the mass of the diquark and the triquark, respectively. $p^2\over 2\mu$ is the kinetic energy of the pentaquark in the center-of-mass system with the relative momentum $p$ and the reduced mass $\mu = \frac{m_{1}m_{2}}{m_{1}+m_{2}}$.

\subsection{Interaction potential}
In addition to the common "Coulomb-plus-linear" potential~\cite{PhysRevLett.34.369,QUIGG1979167,MARTIN1980338,RICHARDSON1979272,KRASEMANN1980397}, Semay and Silvestre-Brac proposed a modified non-relativistic potential~\cite{zpc61.271,fbs20.1}
\begin{align}
    \label{potential}
    V_{q\bar{q}}&=V_{0}+(\vec{s}_{1} \cdot \vec{s}_{2})V_{ss} \notag \\
                &=-\frac{\alpha(1-e^{-r/r_{c}})}{r}+\lambda r^{p}+C \\
                &\  \ \ +(\vec{s}_{1} \cdot \vec{s}_{2})\frac{8\kappa(1-e^{-r/r_{c}})}{3m_{1}}\frac{e^{-r^{2}/r_{0}^{2}}}{r_{0}^{3}} \notag
\end{align}
with $$r_{0}=A(\frac{2m_{1}m_{2}}{m_{1}+m_{2}})^{-B}$$
to describe the mass spectrum of mesons and baryons. The detailed explanation of parameters could be found in original references~\cite{zpc61.271,fbs20.1}.

Due to the two parameters $r_{c}$ and $p$, there are four sets of potentials:
\begin{equation}
    \begin{array}{cc}
        AL1\ \rm {potential}\ :\ p=1,\ r_{c}=0; \\
        AL2\ \rm {potential}\ :\ p=1,\ r_{c}\neq 0; \\
        AP1\ \rm {potential}\ :\ p=\frac{2}{3},\ r_{c}=0; \\
        AP2\ \rm {potential}\ :\ p=\frac{2}{3},\ r_{c}\neq0.\notag
    \end{array}
\end{equation}
When $p=1$, $\lambda r^{p}=\lambda r$ remains the linear confining potential; while $p=\frac{2}{3}$ accounts for the cases with large angular momentum.

For a color singlet state, a strange hidden-charm pentaquark state is formed by a color antitriplet diquark and a color triplet triquark. Accordingly, there are
two types of structures: one with a diquark composed of $cq_{1}$ and a triquark composed of $\bar{c}q_{2}q_{3}$, and another with a diquark composed of $q_{1}q_{2}$ and a triquark composed of $\bar{c}cq_{3}$. One of the $q_{1}$, $q_{2}$, and $q_{3}$ is the $s$ quark and the left two are $u$ or $d$ quark. By assigning the strange quark in different positions within these two structures, we obtain four specific configurations.

Diquarks and triquarks may have spins, there is spin-orbit couplings for cases with orbital angular momentum $L\neq 0$. The spin-orbit coupling potential under the Breit-Fermi approximation is employed as~\cite{prd12.147,prd111.014015} 
\begin{equation}
    \begin{aligned}
    \label{spin-orbit}
     V_{sl}=&V_{so}+V_{ten} \\
     =&\frac{\alpha(1-e^{-r/r_{c}})}{r^{3}}(\frac{1}{m_{1}}+\frac{1}{m_{2}})\times (\frac{\vec{s}_{1}}{m_{1}}+\frac{\vec{s}_{2}}{m_{2}})\cdot \vec{L} \\
            &-\frac{1}{2r}\frac{\partial V_{conf}}{\partial r} (\frac{\vec{s}_{1}}{m_{1}^{2}}+\frac{\vec{s}_{2}}{m_{2}^{2}})\cdot \vec{L} \\
            &+\frac{1}{3m_{1}m_{2}}(\frac{1}{r}\frac{\partial V_{Coul}}{\partial r}-\frac{\partial^{2} V_{Coul}}{\partial r^{2}}) \\
            &\times(\frac{3\vec{s}_{1}\cdot \vec{r} \vec{s}_{2}\cdot \vec{r}}{r^{2}}-\vec{s}_{1} \cdot \vec{s}_{2}).
    \end{aligned}
\end{equation}

In Eq.~\eqref{spin-orbit}, $V_{conf}$ corresponds to the confining potential while $V_{Coul}$ corresponds to the Coulomb potential in Eq.~\eqref{potential}. So the potential in Eq.~\eqref{spin-orbit} has the following form
\begin{equation}
    \begin{aligned}
        \label{SL???}
        V_{sl}=&V_{1}\vec{L}\cdot \vec{s}_{1}+V_{2}\vec{L}\cdot \vec{s}_{2}+V_{t}(\frac{3\vec{s}_{1}\cdot \vec{r} \vec{s}_{2}\cdot \vec{r}}{r^{2}}-\vec{s}_{1} \cdot \vec{s}_{2}) \\
        =&[\frac{\alpha(1-e^{-r/r_{c}})}{m_{1}m_{2}r^{3}}+\frac{1}{m_{1}^{2}}(\frac{\alpha(1-e^{-r/r_{c}})}{2r^{3}}-\frac{\lambda}{2r})]\vec{L}\cdot \vec{s}_{1} \\
        &+[\frac{\alpha(1-e^{-r/r_{c}})}{m_{1}m_{2}r^{3}}+\frac{1}{m_{2}^{2}}(\frac{\alpha(1-e^{-r/r_{c}})}{2r^{3}}-\frac{\lambda}{2r})]\vec{L}\cdot \vec{s}_{2} \\
        &+\frac{\alpha(1-e^{-r/r_{c}})}{m_{1}m_{2}r^{3}}(\frac{3\vec{s}_{1}\cdot \vec{r} \vec{s}_{2}\cdot \vec{r}}{r^{2}}-\vec{s}_{1} \cdot \vec{s}_{2})
    \end{aligned}
\end{equation}
where $\vec{s}_{1}$ and $\vec{s}_{2}$ represent the spin angular momenta of the diquark and triquark respectively, while $\vec{L}$ denotes their orbital angular momentum.

\subsection{Practical calculation}
The Gaussian expansion method~\cite{ppnp51.223} is employed in practical calculations, where the wave functions could be expanded as
\begin{align}
    \label{wave function}
    &\psi _{lm}(\textbf{r} ) = \sum_{n=1}^{n_{max}} c_{nl}\varphi_{nlm}(\textbf{r}), \\
    \label{wave function1}
    &\varphi_{nlm}(\textbf{r}) = N_{nl}r^{l}\textrm{e}^{-r^{2}/r_{n}^{2}}Y_{lm}(\hat{\textbf{r}}).
\end{align}
Three parameters $\{n_{max}, r_{1}, r_{max}\}$ are chosen as variational parameters with $r_{n} = r_{1}a^{n-1}$ where $a=(\frac{r_{n_{max}}}{r_{1}})^{1/(n_{max}-1)}$.

According to the Rayleigh-Ritz variational principle, the solution of the Schrödinger equation is transformed into the following generalized matrix eigenvalue problem:
\begin{equation}
    \sum_{n=1}^{n_{max}} \sum_{n'=1}^{n'_{max}}(H_{nn'}-E_{nl}N_{nn'})c_{nl} = 0
\end{equation}

For the spin-orbit coupling potential, the Wigner symbols are employed for convenience. Denoting the spin of the diquark as $S_{1}$, the spin of the diquark component within the triquark as $S_{d}$, and the spin of the triquark as $S_{2}$, the spin-orbit coupling terms are given by
\begin{align}
   & \left\langle \mathbf{L}\cdot \mathbf{S}_{1} \right\rangle = \sqrt{(2S+1)(2S'+1)} \notag \\
   & \times \sum_{j_{LS_{1}}} \frac{1}{2}[j_{LS_{1}}(j_{LS_{1}}+1)-L(L+1)-S_{1}(S_{1}+1)] \notag \\
   & \times (2j_{LS_{1}}+1) \begin{Bmatrix} L & S_1 & j_{LS_1}\\ S_2 & J & S'\end{Bmatrix} \begin{Bmatrix} L & S_1 & j_{LS_1}\\ S_2 & J & S\end{Bmatrix} .\\
   & \left\langle \mathbf{L}\cdot \mathbf{S}_{2} \right\rangle = \sqrt{(2S+1)(2S'+1)} \notag \\
   & \times \sum_{j_{LS_{2}}} \frac{1}{2}[j_{LS_{2}}(j_{LS_{2}}+1)-L(L+1)-S_{2}(S_{2}+1)] \notag \\
   & \times (2j_{LS_{2}}+1) \begin{Bmatrix} L & S_2 & j_{LS_2}\\ S_1 & J & S'\end{Bmatrix} \begin{Bmatrix} L & S_2 & j_{LS_2}\\ S_1 & J & S\end{Bmatrix}.
\end{align}

The tensor term
\begin{align}
    V_T &= \frac{3(\mathbf{S}_{1}\cdot \mathbf{r}) (\mathbf{S}_{2}\cdot \mathbf{r})}{r^{2}}-(\mathbf{S}_{1} \cdot \mathbf{S}_{2}) \notag \\
    & = 3(\mathbf{S}_{1}\cdot \mathbf{n}) (\mathbf{S}_{2}\cdot \mathbf{n})-(\mathbf{S}_{1}\cdot \mathbf{S}_{2}) \notag \\
    & = 3S_{1i}S_{2j}N_{ij}
\end{align}
where $\mathbf{n} = \mathbf{r}/\mathrm{r}$, and the matrix elements of the tensor operator $N_{ij} = n_{i}n_{j}-\frac{1}{3}\delta_{ij}$ are calculated as follows~\cite{Landau:1991wop}
\begin{align}
    \label{tensor operator}
    \langle N_{ij} \rangle = -\frac{1}{(2L-1)(2L+3)}[L_{i}L_{j}+L_{j}L_{i}-\frac{2}{3}\delta_{ij}L(L+1)].
\end{align}
When the diquark and the triquark are in a S-wave ($L=0$), this term is trivial. For $L=1$, the tensor term yields
\begin{align*}
    V_{T} &=\frac{3(\mathbf{S}_{1}\cdot \mathbf{r}) (\mathbf{S}_{2}\cdot \mathbf{r})}{r^{2}}-(\mathbf{S}_{1} \cdot \mathbf{S}_{2}) \\
    &= -\frac{3}{5} \langle (\mathbf{L} \cdot \mathbf{S_{1}})(\mathbf{L} \cdot \mathbf{S_{2}})+(\mathbf{L} \cdot \mathbf{S_{2}})(\mathbf{L} \cdot \mathbf{S_{1}}) - \frac{4}{3}(\mathbf{S_1}\cdot \mathbf{S_2}) \rangle.
\end{align*}

\section{Numerical results}
As argued in Ref.~\cite{prd111.014015}, the AL potentials are employed to calculate masses of diquarks and triquarks. The predicted masses of diquarks are presented in Table~\ref{diquark-mass}, where spins of diquarks are listed in the second column. The masses for $ud$, $sq$ and $cq$ diquarks are scattered in Refs.~\cite{prd111.014015,arxiv.2602.11744}. The light scalar diquark has a mass $\sim 170$ MeV lower than the light vector diquark.  The predicted mass of scalar $qq$ diquark is about $260$ MeV larger than the predicted one in QCD sum rule, while the predicted mass of scalar $sq$ diquark is almost double as the predicted $460$ MeV in QCD sum rule~\cite{prd76.036004}. The predicted masses of $cq$ diquarks are about $340-410$ MeV larger than the predicted ones in QCD sum rule~\cite{prd87.125018,prd108.054036}. The mass splittings $M_{cs}-M_{cq}$ are about $168$ MeV and $164$ MeV for the scalar and vector diquarks, respectively, which are also much larger than the predicted ones in QCD sum rule~\cite{prd108.054036}. The diquark masses with different flavors may vary from $8$ MeV to $17$ MeV for the two kinds of $AL$ potentials. 

According to spins of triquark and spins of diquark inside, triquark may have three types of spins for each quark flavor. The predicted masses of triquarks are presented in Table~\ref{triquark-mass}, where spins of triquark and diquark are given in the second and third columns, respectively. The same parameters for the calculations of normal mesons and baryons~\cite{zpc61.271,fbs20.1} are employed to calculate the masses of diquarks and triquarks. The mass splittings between the $\bar c qq$ triquark with a scalar diquark and the ones with a vector diquark are about $100-155$ MeV, the mass splittings between the $\bar c sq$ triquark with a scalar diquark and the ones with a vector diquark are about $60-110$ MeV. The mass splittings between the $\bar c cq$ triquark with a scalar diquark and the ones with a vector diquark are about $15-50$ MeV, the mass splittings between the $\bar c cs$ triquark with a scalar diquark and the ones with a vector diquark are about $10-50$ MeV. The triquark masses with each quark flavor may vary from $14$ MeV to $21$ MeV for the two kinds of $AL$ potentials. 
\begin{table}[!htbp]
\centering
\begin{tabular}{cccc}
\toprule
	&spin &AL1 &AL2\\
	\midrule
	$m_{ud}$ & s=0 & 0.666 & 0.674 \\
	&s=1 & 0.834 & 0.842 \\ \hline
	$m_{sq}$ & s=0 & 0.915 & 0.927 \\
	&s=1 & 1.026 &1.038 \\ \hline
	$m_{cs}$ & s=0 & 2.337 & 2.354 \\
	&s=1 & 2.375 & 2.392 \\ \hline
	$m_{cq}$ & s=0 & 2.170 & 2.185 \\
	&s=1 & 2.212 & 2.227 \\
\bottomrule
\end{tabular}
\caption{Masses of diquarks under the AL potential (in GeV)}
\label{diquark-mass}
\end{table}

\begin{table}[!htbp]
\centering
\begin{tabular}{ccccc}
\toprule
	&spin & spin(diquark) &AL1 &AL2\\
	\midrule
	& s=$\frac{1}{2}$ & s=0 & 2.435 & 2.451 \\
	$m_{[\bar{c}qq]}$ & s=$\frac{1}{2}$ & s=1 & 2.537 & 2.551 \\
	& s=$\frac{3}{2}$ & s=1 & 2.590 & 2.605 \\ \hline
	&s=$\frac{1}{2}$& s=0 & 2.644 & 2.661 \\
	$m_{[\bar{c}sq]}$ &s=$\frac{1}{2}$ & s=1 & 2.706 & 2.724 \\
	&s=$\frac{3}{2}$  & s=1 & 2.756 & 2.774 \\ \hline
	&s=$\frac{1}{2}$& s=0 & 3.975 & 3.995 \\
	$m_{[\bar{c}cs]}$ &s=$\frac{1}{2}$ & s=1 & 3.987 & 4.008 \\
	&s=$\frac{3}{2}$  & s=1 & 4.023 & 4.043 \\ \hline
	&s=$\frac{1}{2}$& s=0 & 3.813 & 3.832 \\
	$m_{[\bar{c}cq]}$ &s=$\frac{1}{2}$ & s=1 & 3.829 & 3.847 \\
	&s=$\frac{3}{2}$  & s=1 & 3.865 & 3.884 \\		
\bottomrule
\end{tabular}
\caption{Masses of triquarks under the AL potential (in GeV)}
\label{triquark-mass}
\end{table}

For S- and P-wave excitations between the diquark and the triquark, the AL potentials may be preferred to calculate masses of pentaquarks~\cite{prd111.014015}. Therefore, there are four different combinations: AL1-AL1, AL1-AL2, AL2-AL1, and AL2-AL2 for the calculation of masses of pentaquarks. In these four types of combinations, the first term indicates the AL potentials used for the interaction inside diquark and triquark, while the second term represents the AL potentials employed for the interaction between the diquark and the triquark in a pentaquark. The same parameters $\alpha$ and $\lambda$ in the calculation of tetraquark states are employed to calculate masses of pentaquark states~\cite{zpc61.271,fbs20.1,prd111.014015}. These parameters are presented in Table~\ref{pentaquark-parameter}.

\begin{table}[!htbp]
\centering
\begin{tabular}{ccccc}
\toprule
	Parameters &AL1-AL1 &AL1-AL2 &AL2-AL1 &AL2-AL2\\
	\midrule
	$\alpha$ & 0.3910 & 0.4205 & 0.4290 & 0.4600 \\
	$\kappa$& 1.8609 & 1.8475 & 1.8609 & 1.8475 \\ 
	$\lambda(\mathrm{GeV}^2)$ & 0.1941 & 0.1916 & 0.1957 & 0.1932 \\
	$C(\mathrm{GeV})$& -0.8321 & -0.8182 & -0.8321 & -0.8182 \\ 
	$B$ & 0.2204 & 0.2132 & 0.2204 & 0.2132 \\
	$A(\mathrm{GeV}^{B-1})$ & 1.6653 & 1.6560 & 1.6653 & 1.6560 \\ 
	$r_{c}(\mathrm{GeV}^{-1})$ & 0 & 0.1844 & 0 & 0.1844 \\
			
\bottomrule
\end{tabular}
\caption{Parameters of different combinations}
\label{pentaquark-parameter}
\end{table}

In terms of the predicted masses for diquarks and triquarks, we have obtained mass spectrum of the pentaquark states within four different types of combinations (AL1-AL1, AL2-AL1, AL1-AL2, and AL2-AL2). According to the four specific flavor configurations, the numerical results are presented in four tables: mass spectrum for strange hidden-charm pentaquark states in $[cs][\bar{c}qq]$ configuration is given in Table~\ref{pentaquark-mass1}. Table~\ref{pentaquark-mass2} lists the results for pentaquark states in $[cq][\bar{c}sq]$ configuration. The masses in $[qq][\bar{c}cs]$ configuration are presented in Table \ref{pentaquark-mass3}. Finally, the spectrum in $[sq][\bar{c}cq]$ configuration is shown in Table~\ref{pentaquark-mass4}. In the second column in these four tables, the convention $|S_1; S_d, S_2; S, L \rangle_J$ for the pentaquark state is employed. 

\begin{table*}[!htbp]
\centering
\begin{tabular}{cccccc}
\toprule

        $J^P$ & $| S_1; S_d, S_2; S, L \rangle_J$  &AL1-AL1 &AL2-AL1 &AL1-AL2 &AL2-AL2 \\ \midrule
        $1/2^-$ & $| 0; 0, 1/2; 1/2, 0 \rangle_{1/2}$ & 4234.66 & 4235.72 & 4234.94 & 4236.34 \\ 
        $1/2^-$ & $| 0; 1, 1/2; 1/2, 0 \rangle_{1/2}$ & 4330.23 & 4329.22 & 4330.64 & 4330.01 \\ 
        $3/2^-$ & $| 0; 1, 3/2; 3/2, 0 \rangle_{3/2}$ & 4380.04 & 4379.86 & 4380.51 & 4380.74 \\ 
        $1/2^-$ & $| 1; 0, 1/2; 1/2, 0 \rangle_{1/2}$ & 4197.92 & 4197.69 & 4202.66 & 4203.19 \\ 
        $3/2^-$ & $| 1; 0, 1/2; 3/2, 0 \rangle_{3/2}$ & 4303.33 & 4304.97 & 4301.95 & 4303.75 \\ 
        $1/2^-$ & $| 1; 1, 1/2; 1/2, 0 \rangle_{1/2}$ & 4294.93 & 4292.60 & 4299.81 & 4298.28 \\ 
        $3/2^-$ & $| 1; 1, 1/2; 3/2, 0 \rangle_{3/2}$ & 4398.27 & 4397.81 & 4396.95 & 4396.73 \\ 
        $1/2^-$ & $| 1; 1, 3/2; 1/2, 0 \rangle_{1/2}$ & 4228.93 & 4225.85 & 4242.74 & 4241.10 \\ 
        $3/2^-$ & $| 1; 1, 3/2; 3/2, 0 \rangle_{3/2}$ & 4345.47 & 4343.97 & 4350.41 & 4349.75 \\ 
        $5/2^-$ & $| 1; 1, 3/2; 5/2, 0 \rangle_{5/2}$ & 4507.25 & 4509.01 & 4503.43 & 4505.06 \\ 

        $1/2^+$ & $| 0; 0, 1/2; 1/2, 1 \rangle_{1/2}$ & 4641.82 & 4649.17 & 4640.42 & 4646.91 \\ 
        $3/2^+$ & $| 0; 0, 1/2; 1/2, 1 \rangle_{3/2}$ & 4606.00 & 4618.38 & 4608.50 & 4620.16 \\ 

        $1/2^+$ & $| 0; 1, 1/2; 1/2, 1 \rangle_{1/2}$ & 4734.29 & 4740.01 & 4732.75 & 4737.63 \\ 
        $3/2^+$ & $| 0; 1, 1/2; 1/2, 1 \rangle_{3/2}$ & 4702.90 & 4713.30 & 4705.21 & 4714.88 \\ 

        $1/2^+$ & $| 0; 1, 3/2; 3/2, 1 \rangle_{1/2}$ & 4640.63 & 4896.62 & 4701.22 & 4879.35 \\ 
        $3/2^+$ & $| 0; 1, 3/2; 3/2, 1 \rangle_{3/2}$ & 4782.66 & 4789.38 & 4781.05 & 4786.92 \\ 
        $5/2^+$ & $| 0; 1, 3/2; 3/2, 1 \rangle_{5/2}$ & 4726.98 & 4740.47 & 4730.76 & 4743.61 \\ 

        $1/2^+$ & $| 1; 0, 1/2; 1/2, 1 \rangle_{1/2}$ & 4652.69 & 4659.36 & 4650.52 & 4656.36 \\ 
        $3/2^+$ & $| 1; 0, 1/2; 1/2, 1 \rangle_{3/2}$ & 4622.26 & 4634.57 & 4624.33 & 4635.89\\ 

        $1/2^+$ & $| 1; 0, 1/2; 3/2, 1 \rangle_{1/2}$  & 4507.43 & 4545.96 & 4766.21 & 4728.53 \\
        $3/2^+$ & $| 1; 0, 1/2; 3/2, 1 \rangle_{3/2}$ & 4726.37 & 4737.47 & 4729.97 & 4740.39 \\ 
        $5/2^+$ & $| 1; 0, 1/2; 3/2, 1 \rangle_{5/2}$ & 4608.64 & 4623.09 & 4612.11 & 4625.89 \\ 

        $1/2^+$ & $| 1; 1, 1/2; 1/2, 1 \rangle_{1/2}$ & 4746.94 & 4751.86 & 4744.62 & 4748.71 \\ 
        $3/2^+$ & $| 1; 1, 1/2; 1/2, 1 \rangle_{3/2}$ & 4718.63 & 4729.00 & 4720.54 & 4730.15 \\ 

        $1/2^+$ & $| 1; 1, 1/2; 3/2, 1 \rangle_{1/2}$ & 4624.71 & 4496.50 & 4874.21 & 4830.29 \\ 
        $3/2^+$ & $| 1; 1, 1/2; 3/2, 1 \rangle_{3/2}$ & 4819.07 & 4828.39 & 4822.47 & 4831.12 \\ 
        $5/2^+$ & $| 1; 1, 1/2; 3/2, 1 \rangle_{5/2}$ & 4706.79 & 4719.20 & 4710.07 & 4721.81 \\ 

        $1/2^+$ & $| 1; 1, 3/2; 1/2, 1 \rangle_{1/2}$ & 4754.23 & 4759.09 & 4750.99 & 4755.11 \\ 
        $3/2^+$ & $| 1; 1, 3/2; 1/2, 1 \rangle_{3/2}$ & 4737.56 & 4748.66 & 4738.95 & 4749.27 \\ 

        $1/2^+$ & $| 1; 1, 3/2; 3/2, 1 \rangle_{1/2}$ & 4876.43 & 4835.60 & 4698.35 & 4641.82 \\ 
        $3/2^+$ & $| 1; 1, 3/2; 3/2, 1 \rangle_{3/2}$ & 4859.89 & 4871.50 & 4864.83 & 4875.84 \\ 
        $5/2^+$ & $| 1; 1, 3/2; 3/2, 1 \rangle_{5/2}$ & 4729.74 & 4742.53 & 4731.91 & 4743.96 \\ 

        $3/2^+$ & $| 1; 1, 3/2; 5/2, 1 \rangle_{3/2}$ & 4871.67 & 4841.58 & 4676.74 & 4552.30 \\ 
        $5/2^+$ & $| 1; 1, 3/2; 5/2, 1 \rangle_{5/2}$ & 4916.18 & 4928.58 & 4923.47 & 4935.36 \\ 
        $7/2^+$ & $| 1; 1, 3/2; 5/2, 1 \rangle_{7/2}$ & 5078.16 & 5092.66 & 5086.59 & 5100.62 \\ 
\bottomrule
\end{tabular}
\caption{Masses of pentaquarks ($[cs][\bar{c}qq]$) under the AL potential (in MeV)}
\label{pentaquark-mass1}
\end{table*}

\begin{table*}[!htbp]
\centering
\begin{tabular}{cccccc}
\toprule

        $J^P$ & $| S_1; S_d, S_2; S, L \rangle_J$  &AL1-AL1 &AL2-AL1 &AL1-AL2 &AL2-AL2 \\ \midrule
        $1/2^-$ & $| 0; 0, 1/2; 1/2, 0 \rangle_{1/2}$ & 4276.84 & 4276.98 & 4277.12 & 4277.60 \\
        $1/2^-$ & $| 0; 1, 1/2; 1/2, 0 \rangle_{1/2}$ & 4335.48 & 4336.47 & 4335.83 & 4337.18 \\
        $3/2^-$ & $| 0; 1, 3/2; 3/2, 0 \rangle_{3/2}$ & 4382.86 & 4383.77 & 4383.26 & 4386.54 \\
        $1/2^-$ & $| 1; 0, 1/2; 1/2, 0 \rangle_{1/2}$ & 4244.07 & 4242.89 & 4248.79 & 4248.37 \\
        $3/2^-$ & $| 1; 0, 1/2; 3/2, 0 \rangle_{3/2}$ & 4348.47 & 4349.17 & 4347.11 & 4347.98 \\
        $1/2^-$ & $| 1; 1, 1/2; 1/2, 0 \rangle_{1/2}$ & 4303.58 & 4303.25 & 4308.36 & 4308.82 \\
        $3/2^-$ & $| 1; 1, 1/2; 3/2, 0 \rangle_{3/2}$ & 4406.71 & 4408.24 & 4405.40 & 4407.12 \\
        $1/2^-$ & $| 1; 1, 3/2; 1/2, 0 \rangle_{1/2}$ & 4235.19 & 4233.24 & 4248.79 & 4248.24 \\
        $3/2^-$ & $| 1; 1, 3/2; 3/2, 0 \rangle_{3/2}$ & 4351.64 & 4351.23 & 4356.47 & 4356.87 \\
        $5/2^-$ & $| 1; 1, 3/2; 5/2, 0 \rangle_{5/2}$ & 4513.21 & 4516.01 & 4509.40 & 4512.05 \\

        $1/2^+$ & $| 0; 0, 1/2; 1/2, 1 \rangle_{1/2}$ & 4679.27 & 4686.08 & 4677.95 & 4683.93 \\
        $3/2^+$ & $| 0; 0, 1/2; 1/2, 1 \rangle_{3/2}$ & 4651.12 & 4662.44 & 4653.57 & 4664.16 \\

        $1/2^+$ & $| 0; 1, 1/2; 1/2, 1 \rangle_{1/2}$ & 4782.27 & 4790.61 & 4781.23 & 4788.32 \\
        $3/2^+$ & $| 0; 1, 1/2; 1/2, 1 \rangle_{3/2}$ & 4758.21 & 4770.41 & 4760.48 & 4771.95 \\

        $1/2^+$ & $| 0; 1, 3/2; 3/2, 1 \rangle_{1/2}$ & 4726.36 & 4442.32 & 4747.30 & 4587.42 \\
        $3/2^+$ & $| 0; 1, 3/2; 3/2, 1 \rangle_{3/2}$ & 4782.70 & 4790.61 & 4781.23 & 4788.31 \\
        $5/2^+$ & $| 0; 1, 3/2; 3/2, 1 \rangle_{5/2}$ & 4736.05 & 4750.32 & 4739.84 & 4753.47 \\

        $1/2^+$ & $| 1; 0, 1/2; 1/2, 1 \rangle_{1/2}$ & 4698.61 & 4704.38 & 4696.43 & 4701.38 \\
        $3/2^+$ & $| 1; 0, 1/2; 1/2, 1 \rangle_{3/2}$ & 4667.73 & 4679.11 & 4669.79 & 4680.42 \\

        $1/2^+$ & $| 1; 0, 1/2; 3/2, 1 \rangle_{1/2}$ & 4562.77 & 4414.29 & 4817.78 & 4777.68 \\
        $3/2^+$ & $| 1; 0, 1/2; 3/2, 1 \rangle_{3/2}$ & 4771.64 & 4781.83 & 4775.20 & 4784.71 \\
        $5/2^+$ & $| 1; 0, 1/2; 3/2, 1 \rangle_{5/2}$ & 4653.72 & 4667.23 & 4657.17 & 4670.11 \\

        $1/2^+$ & $| 1; 1, 1/2; 1/2, 1 \rangle_{1/2}$ & 4756.66 & 4763.43 & 4754.40 & 4760.35 \\
        $3/2^+$ & $| 1; 1, 1/2; 1/2, 1 \rangle_{3/2}$ & 4726.79 & 4739.09 & 4728.76 & 4740.31 \\

        $1/2^+$ & $| 1; 1, 1/2; 3/2, 1 \rangle_{1/2}$ & 4633.91 & 4505.95 & 4884.19 & 4842.47 \\
        $3/2^+$ & $| 1; 1, 1/2; 3/2, 1 \rangle_{3/2}$ & 4828.77 & 4839.90 & 4832.22 & 4842.68 \\
        $5/2^+$ & $| 1; 1, 1/2; 3/2, 1 \rangle_{5/2}$ & 4713.60 & 4728.01 & 4716.95 & 4730.69 \\

        $1/2^+$ & $| 1; 1, 3/2; 1/2, 1 \rangle_{1/2}$ & 4761.69 & 4767.49 & 4758.53 & 4763.59 \\
        $3/2^+$ & $| 1; 1, 3/2; 1/2, 1 \rangle_{3/2}$ & 4743.34 & 4755.45 & 4744.80 & 4756.13 \\

        $1/2^+$ & $| 1; 1, 3/2; 3/2, 1 \rangle_{1/2}$ & 4884.30 & 4844.42 & 4705.70 & 4649.66 \\
        $3/2^+$ & $| 1; 1, 3/2; 3/2, 1 \rangle_{3/2}$ & 4867.38 & 4879.89 & 4872.39 & 4884.29 \\
        $5/2^+$ & $| 1; 1, 3/2; 3/2, 1 \rangle_{5/2}$ & 4734.05 & 4747.90 & 4736.31 & 4749.43 \\

        $3/2^+$ & $| 1; 1, 3/2; 5/2, 1 \rangle_{3/2}$ & 4880.72 & 4851.23 & 4682.87 & 4556.61 \\
        $5/2^+$ & $| 1; 1, 3/2; 5/2, 1 \rangle_{5/2}$ & 4923.72 & 4937.00 & 4931.07 & 4943.84 \\
        $7/2^+$ & $| 1; 1, 3/2; 5/2, 1 \rangle_{7/2}$ & 5079.87 & 5095.51 & 5088.37 & 5103.55 \\
\bottomrule
\end{tabular}
\caption{Masses of pentaquarks ($[cq][\bar{c}sq]$)  under the AL potential (in MeV)}
\label{pentaquark-mass2}
\end{table*}

\begin{table*}[!htbp]
\centering
\begin{tabular}{cccccc}
\toprule
        $J^P$ & $| S_1; S_d, S_2; S, L \rangle_J$  &AL1-AL1 &AL2-AL1 &AL1-AL2 &AL2-AL2 \\ \midrule
        $1/2^-$ & $| 0; 0, 1/2; 1/2, 0 \rangle_{1/2}$ & 4343.75 & 4344.65 & 4344.27 & 4344.63 \\

        $1/2^-$ & $| 0; 1, 1/2; 1/2, 0 \rangle_{1/2}$ & 4355.60 & 4357.49 & 4356.12 & 4357.47 \\

        $3/2^-$ & $| 0; 1, 3/2; 3/2, 0 \rangle_{3/2}$ & 4391.17 & 4392.06 & 4391.69 & 4392.04 \\

        $1/2^-$ & $| 1; 0, 1/2; 1/2, 0 \rangle_{1/2}$ & 4378.56 & 4378.26 & 4380.36 & 4379.80 \\
        $3/2^-$ & $| 1; 0, 1/2; 3/2, 0 \rangle_{3/2}$ & 4481.15 & 4482.06 & 4480.42 & 4482.08 \\

        $1/2^-$ & $| 1; 1, 1/2; 1/2, 0 \rangle_{1/2}$ & 4390.57 & 4391.27 & 4392.36 & 4392.80 \\
        $3/2^-$ & $| 1; 1, 1/2; 3/2, 0 \rangle_{3/2}$ & 4492.91 & 4494.79 & 4492.17 & 4493.55 \\

        $1/2^-$ & $| 1; 1, 3/2; 1/2, 0 \rangle_{1/2}$ & 4307.18 & 4305.84 & 4313.54 & 4312.24 \\
        $3/2^-$ & $| 1; 1, 3/2; 3/2, 0 \rangle_{3/2}$ & 4426.61 & 4426.30 & 4428.39 & 4427.82 \\
        $5/2^-$ & $| 1; 1, 3/2; 5/2, 0 \rangle_{5/2}$ & 4585.88 & 4587.60 & 4584.48 & 4585.57 \\

        $1/2^+$ & $| 0; 0, 1/2; 1/2, 1 \rangle_{1/2}$ & 4772.22& 4775.25 & 4776.72 & 4779.51 \\
        $3/2^+$ & $| 0; 0, 1/2; 1/2, 1 \rangle_{3/2}$ & 4756.71 & 4762.48 & 4764.23 & 4639.36 \\

        $1/2^+$ & $| 0; 1, 1/2; 1/2, 1 \rangle_{1/2}$ & 4783.98 & 4787.98 & 4788.46 & 4792.23 \\
        $3/2^+$ & $| 0; 1, 1/2; 1/2, 1 \rangle_{3/2}$ & 4768.58 & 4775.34 & 4776.11 & 4782.32 \\

        $1/2^+$ & $| 0; 1, 3/2; 3/2, 1 \rangle_{1/2}$ & 4823.15 & 4821.48 & 4826.69 & 4824.85 \\
        $3/2^+$ & $| 0; 1, 3/2; 3/2, 1 \rangle_{3/2}$ & 4819.32 & 4822.27 & 4823.67 & 4826.48 \\
        $5/2^+$ & $| 0; 1, 3/2; 3/2, 1 \rangle_{5/2}$ & 4792.25 & 4799.52 & 4801.08 & 4807.60 \\

        $1/2^+$ & $| 1; 0, 1/2; 1/2, 1 \rangle_{1/2}$ & 5013.71 & 5007.56 & 5014.93 & 5008.22 \\
        $3/2^+$ & $| 1; 0, 1/2; 1/2, 1 \rangle_{3/2}$ & 4728.20 & 4739.29 & 4734.72 & 4745.26 \\

        $1/2^+$ & $| 1; 0, 1/2; 3/2, 1 \rangle_{1/2}$ & 5419.35 & 5334.67 & 5204.79 & 5135.25 \\
        $3/2^+$ & $| 1; 0, 1/2; 3/2, 1 \rangle_{3/2}$ & 5103.63 & 5102.70 & 5110.67 & 5109.14 \\
        $5/2^+$ & $| 1; 0, 1/2; 3/2, 1 \rangle_{5/2}$ & 5003.36 &  5030.09 & 5016.73 & 5043.31 \\

        $1/2^+$ & $| 1; 0, 1/2; 1/2, 1 \rangle_{1/2}$ & 5025.55 & 5020.39 & 5026.77 & 5021.05 \\
        $3/2^+$ & $| 1; 0, 1/2; 1/2, 1 \rangle_{3/2}$ & 4740.08 & 4752.16 & 4746.60 & 4758.13 \\

        $1/2^+$ & $| 1; 1, 1/2; 3/2, 1 \rangle_{1/2}$ & 5433.04 & 5348.81 & 5217.65 & 5149.50 \\
        $3/2^+$ & $| 1; 1, 1/2; 3/2, 1 \rangle_{3/2}$ & 5115.24 & 5115.27 & 5122.26 & 5121.70 \\
        $5/2^+$ & $| 1; 1, 1/2; 3/2, 1 \rangle_{5/2}$ & 5015.35 & 5043.07 & 5028.71 & 5056.29 \\

        $1/2^+$ & $| 1; 1, 3/2; 1/2, 1 \rangle_{1/2}$ & 5010.61 & 5003.40 & 5011.12 & 5003.49 \\
        $3/2^+$ & $| 1; 1, 3/2; 1/2, 1 \rangle_{3/2}$ & 4751.85 & 4762.87 & 4757.80 & 4768.39 \\

        $1/2^+$ & $| 1; 1, 3/2; 3/2, 1 \rangle_{1/2}$ & 5204.65 & 5127.61 & 4806.03 & 4525.21 \\
        $3/2^+$ & $| 1; 1, 3/2; 3/2, 1 \rangle_{3/2}$ & 5141.11 & 5141.92 & 5150.18 & 5150.45 \\
        $5/2^+$ & $| 1; 1, 3/2; 3/2, 1 \rangle_{5/2}$ & 5039.17 & 5065.77 & 5052.40 & 5078.83 \\

        $3/2^+$ & $| 1; 1, 3/2; 5/2, 1 \rangle_{3/2}$ & 5272.12 & 5123.85 & 5739.19 & 5617.82 \\
        $5/2^+$ & $| 1; 1, 3/2; 5/2, 1 \rangle_{5/2}$ & 5210.76 &  5211.95 & 5222.41 & 5223.15 \\
        $7/2^+$ & $| 1; 1, 3/2; 5/2, 1 \rangle_{7/2}$ & 5317.63 &  5346.10 & 5324.15 & 5352.47 \\
\bottomrule
\end{tabular}
\caption{Masses of pentaquarks ($[qq][\bar{c}cs]$) under the AL potential (in MeV)}
\label{pentaquark-mass3}
\end{table*}

\begin{table*}[!htbp]
\centering
\begin{tabular}{cccccc}
\toprule
        $J^P$ & $| S_1; S_d, S_2; S, L \rangle_J$  &AL1-AL1 &AL2-AL1 &AL1-AL2 &AL2-AL2 \\ \midrule
        $1/2^-$ & $| 0; 0, 1/2; 1/2, 0 \rangle_{1/2}$ & 4345.74 & 4347.86 & 4345.47 & 4347.20 \\

        $1/2^-$ & $| 0; 1, 1/2; 1/2, 0 \rangle_{1/2}$ & 4361.48 & 4362.60 & 4361.20 & 4361.94 \\

        $3/2^-$ & $| 0; 1, 3/2; 3/2, 0 \rangle_{3/2}$ & 4396.89 & 4398.98 & 4396.61 & 4398.32 \\

        $1/2^-$ & $| 1; 0, 1/2; 1/2, 0 \rangle_{1/2}$ & 4372.63 & 4372.58 & 4374.68 & 4374.55 \\
        $3/2^-$ & $| 1; 0, 1/2; 3/2, 0 \rangle_{3/2}$ & 4459.27 & 4461.57 & 4457.99 & 4459.89 \\

        $1/2^-$ & $| 1; 1, 1/2; 1/2, 0 \rangle_{1/2}$ & 4356.68 & 4357.62 & 4358.73 & 4359.61 \\
        $3/2^-$ & $| 1; 1, 1/2; 3/2, 0 \rangle_{3/2}$ & 4474.88 & 4476.20 & 4473.60 & 4474.51 \\

        $1/2^-$ & $| 1; 1, 3/2; 1/2, 0 \rangle_{1/2}$ & 4290.12 & 4289.92 & 4297.74 & 4297.84 \\
        $3/2^-$ & $| 1; 1, 3/2; 3/2, 0 \rangle_{3/2}$ & 4408.53 & 4409.46 & 4410.57 & 4411.42 \\
        $5/2^-$ & $| 1; 1, 3/2; 5/2, 0 \rangle_{5/2}$ & 4568.01 & 4571.25 & 4565.68 & 4568.33 \\

        $1/2^+$ & $| 0; 0, 1/2; 1/2, 1 \rangle_{1/2}$ & 4757.37& 4754.35 & 4759.89 & 4765.29 \\
        $3/2^+$ & $| 0; 0, 1/2; 1/2, 1 \rangle_{3/2}$ & 4743.38 & 4743.21 & 4749.18 & 4757.26 \\

        $1/2^+$ & $| 0; 1, 1/2; 1/2, 1 \rangle_{1/2}$ & 4772.98 & 4777.74 & 4775.46 & 4779.89 \\
        $3/2^+$ & $| 0; 1, 1/2; 1/2, 1 \rangle_{3/2}$ & 4759.15 & 4766.85 & 4764.94 & 4772.02 \\

        $1/2^+$ & $| 0; 1, 3/2; 3/2, 1 \rangle_{1/2}$ & 4807.66 & 4807.02 & 4809.51 & 4809.10 \\
        $3/2^+$ & $| 0; 1, 3/2; 3/2, 1 \rangle_{3/2}$ & 4808.14 & 4813.82 & 4810.51 & 4815.93 \\
        $5/2^+$ & $| 0; 1, 3/2; 3/2, 1 \rangle_{5/2}$ & 4783.66 & 4793.66 & 4790.57 & 4800.02 \\

        $1/2^+$ & $| 1; 0, 1/2; 1/2, 1 \rangle_{1/2}$ & 4913.79 & 4912.93 & 4914.19 & 4912.61 \\
        $3/2^+$ & $| 1; 0, 1/2; 1/2, 1 \rangle_{3/2}$ & 4741.23 & 4753.40 & 4746.55 & 4758.10 \\

        $1/2^+$ & $| 1; 0, 1/2; 3/2, 1 \rangle_{1/2}$ & 5304.86 & 5205.16 & 5066.50 & 5018.11 \\
        $3/2^+$ & $| 1; 0, 1/2; 3/2, 1 \rangle_{3/2}$ & 4995.47& 4999.13 & 5001.53 & 5004.57 \\
        $5/2^+$ & $| 1; 0, 1/2; 3/2, 1 \rangle_{5/2}$ & 4940.13 &  4961.25 & 4951.15 &  4971.92 \\

        $1/2^+$ & $| 1; 1, 1/2; 1/2, 1 \rangle_{1/2}$ & 4929.49 & 4927.66 & 4929.88 & 4927.34 \\
        $3/2^+$ & $| 1; 1, 1/2; 1/2, 1 \rangle_{3/2}$ & 4757.01 & 4768.19 & 4762.32 & 4772.88 \\

        $1/2^+$ & $| 1; 1, 1/2; 3/2, 1 \rangle_{1/2}$ & 5324.10 & 5222.01 & 5083.38 & 5034.04 \\
        $3/2^+$ & $| 1; 1, 1/2; 3/2, 1 \rangle_{3/2}$ & 5010.90 & 5013.59 & 5016.94 & 5019.02 \\
        $5/2^+$ & $| 1; 1, 1/2; 3/2, 1 \rangle_{5/2}$ & 4956.01 & 4976.12 & 4967.02 & 4986.79 \\

        $1/2^+$ & $| 1; 1, 3/2; 1/2, 1 \rangle_{1/2}$ & 4918.37 & 4916.62 & 4917.99 & 4915.61 \\
        $3/2^+$ & $| 1; 1, 3/2; 1/2, 1 \rangle_{3/2}$ & 4766.12 & 4778.18 & 4770.99 & 4782.41 \\

        $1/2^+$ & $| 1; 1, 3/2; 3/2, 1 \rangle_{1/2}$ & 5075.05 & 5017.59 & 4804.42 & 4650.99 \\
        $3/2^+$ & $| 1; 1, 3/2; 3/2, 1 \rangle_{3/2}$ & 5037.31 & 5042.55 & 5045.21 &  5049.89 \\
        $5/2^+$ & $| 1; 1, 3/2; 3/2, 1 \rangle_{5/2}$ & 4975.14 & 4995.99 & 4985.79 & 5006.27 \\

        $3/2^+$ & $| 1; 1, 3/2; 5/2, 1 \rangle_{3/2}$ & 5137.18 & 5053.03 & 5558.11 & 5446.86 \\
        $5/2^+$ & $| 1; 1, 3/2; 5/2, 1 \rangle_{5/2}$ & 5101.06 &  5106.73 & 5111.40 & 5116.61 \\
        $7/2^+$ & $| 1; 1, 3/2; 5/2, 1 \rangle_{7/2}$ & 5154.01 &  5187.06 & 5165.66 & 5198.71 \\
\bottomrule
\end{tabular}
\caption{Masses of pentaquarks ($[sq][\bar{c}cq]$)  under the AL potential (in MeV)}
\label{pentaquark-mass4}
\end{table*}

Obviously, the pentaquark states make a rich spectroscopy of baryons in the diquark-triquark structure. There are 10 negative-parity pentaquark states with $L=0$ and 25 positive-parity pentaquark states with $L=1$ for each type of flavor configurations. The mass gap between the S-wave and P-wave pentaquark states is about $350-570$ MeV. 

The masses of constituent diquarks have 8-17 MeV difference and the masses of constituent triquarks have $14-21$ MeV difference between the AL1 and AL2 potentials. The mass difference of the same S-wave pentaquark state among the four combinations is several MeV, and it is only $1\to 2$ MeV for lowest ground states. As well known, a heavier reduced mass will suppress the kinetic energy of the system and will lead naturally to a spatial contraction of the wave function. This contraction will finally drive the system deeper into the color-Coulomb potential, and yield additional attraction which offsets the increase of constituent masses.

The mass difference of the same P-wave pentaquark state among the four combinations is up to $44$ MeV, which implies that this offset has small affect on the P-wave excitations. 

This offset effect could also be assumed as a dynamical compensation mechanism within the variational principle, which indicates the robust self-consistency of the Gaussian expansion method (GEM) employed in multiquark systems.

From Table~\ref{pentaquark-mass1} to Table~\ref{pentaquark-mass4}, all predicted masses for positive-parity ($J^+$) pentaquark states with $L=1$ are above $4600$ MeV. In contrast, the masses of negative-parity pentaquark states with $L=0$ are distributed between $4200$ MeV and $4590$ MeV.

For a $J^{P}=1/2^{-}$ pentaquark state at S-wave ($L=0$), the numerical results indicate that the $| 0; 0, 1/2; 1/2, 0 \rangle_{1/2}$ pentaquark state with $J^{P}=1/2^{-}$ has mass about $4344$ MeV and $4345$ MeV in configurations $[qq][\bar{c}cs]$ and $[sq][\bar{c}cq]$, respectively, which is slightly larger than the experimental value. In configuration $[cs][\bar{c}qq]$, the state $|0; 1, 1/2; 1/2, 0\rangle_{1/2}$ with $J^{P}=1/2^{-}$ has mass about $4330$ MeV, slightly smaller than the experimental value. In configuration $[cq][\bar{c}sq]$, the pentaquark state $|0; 1, 1/2; 1/2, 0\rangle_{1/2}$ with $J^{P}=1/2^{-}$ has mass about $4336$ MeV. Therefore, it is very possible that $P_{c\bar c s}(4338)^{0}$ is the $|0; 1, 1/2; 1/2, 0\rangle_{1/2}$ pentaquark state in configuration $[cq][\bar{c}sq]$. Of course, other three configurations can not be ruled out under uncertainties of numerical results.

When the $|0; 1,1/2; 1/2, 0\rangle_{1/2}$ pentaquark state is in the $[cq][\bar{c}sq]$ configuration, it provides a natural dynamical explanation for its extremely narrow decay width into $J/\psi\Lambda$. In this configuration, the charm quark resides in a tightly bound scalar $cq$ diquark, while the anti-charm quark is confined within the spatially separated triquark cluster. As emphasized in the dynamical diquark picture, these diquark and triquark clusters can be considered as well-separated components and the spatial separation between the $c$ and $\bar{c}$ quarks is expected to be larger than the typical compact size of the $J/\psi$ wave function \cite{LEBED2015454}. This spatial segregation significantly suppresses the short-range overlap required to form a $c\bar{c}$ bound state. Furthermore, decaying into the $J/\psi\Lambda$ final state requires breaking these pre-existing strongly bound clusters and undergoing an extensive rearrangement of the internal color-spin configurations to form color-singlets ~\cite{MAIANI2015289, PhysRevLett.113.112001}. This complex color-spin and spatial recombination inherently suppresses the transition amplitude. Therefore, the pentaquark state with the interior structure assumed in the calculations of masses exhibits a narrow decay width.

From Table~\ref{pentaquark-mass4}, the $|1; 0, 1/2; 3/2, 0\rangle_{3/2}$ pentaquark state has mass around 4459 MeV in configuration $[sq][\bar{c}cq]$ for four potentials combinations. When $P_{c\bar cs}(4459)^{0}$ is assumed as the $|1; 0, 1/2; 3/2, 0\rangle_{3/2}$ pentaquark state in configuration $[sq][\bar{c}cq]$, the predicted mass is consistent with experimental data. The $J^{P}$ quantum numbers of $P_{c\bar cs}(4459)^{0}$ could be read out with $J^P={3\over 2}^-$ though they have not yet been determined in experiment. 

The $|1; 0, 1/2; 3/2, 0\rangle_{3/2}$ in configuration $[sq][\bar{c}cq]$ contains a vector diquark which possesses higher color-magnetic energy compared to a scalar diquark, which is the origin that $P_{c\bar cs}(4459)^{0}$ is about $120$ MeV heavier than $P_{c\bar c s}(4338)^{0}$. The hyperfine mass splitting is close to the mass splitting between $D$ and $D^{*}$ mesons. Moreover, in configuration $[sq][\bar{c}cq]$, the $c$ and $\bar{c}$ are within the same tightly bound subsystem, and lead to a larger wave function overlap and a higher probability of recombination into $J/\psi$ compared to the case of $P_{c\bar c s}(4338)^{0}$. This configuration accounts for an observed broader decay width of $P_{c\bar cs}(4459)^{0}$ ($17.3 \pm 6.5 ^{+8.0}_{-5.7}$ MeV) in comparison to that of $P_{c\bar c s}(4338)^{0}$.

Our predicted masses of these two pentaquark candidates in the diquark-triquark picture are near the $\Xi_{c}D$ and $\Xi_{c}D^{*}$ thresholds, but are about 18 MeV below the $\Xi_{c}D^{*}$ threshold. They cannot be $\Xi_{c}D$ and $\Xi_{c}D^{*}$ molecular states since boson exchange interaction typically cannot produce such a deep binding energy. In contrast, there is no such a problem for pentaquark states in the diquark-triquark configuration.

\section{Summary}
Within a non-relativistic constituent quark model, strange hidden-charm pentaquark states from S-wave to P-wave excitation have been studied. In this model, a pentaquark state consists of a diquark and a triquark with color, and the triquark consists of a diquark and an antiquark. The Semay and Silvestre-Brac potentials are employed to account for interior interactions in diquark, triquark and pentaquark states, where all the color-magnetic interactions, spin-orbit coupling, and tensor forces have been taken into account. In practical calculation, the same parameters in calculations of tetraquark states are employed for calculations of pentaquark states, and the Gaussian Expansion Method (GEM) is employed in four quark flavor configurations in four types of potentials combinations. Masses of diquarks and triquarks are calculated first, and masses of strange hidden-charm pentaquark states are subsequently obtained. 

The predicted masses of scalar $cq$ diquarks are about $2170-2185$ MeV, and the vector ones are about $2212-2227$ MeV. They are $340-410$ MeV larger than the predicted masses in QCD sum rule. The mass splittings $M_{cs}-M_{cq}$ are about $168$ MeV and $164$ MeV for the scalar and vector diquarks, respectively. The predicted masses of triquarks vary from $2435$ MeV to $4023$ MeV for different quark flavors with anti-charm quark. The mass splittings between the triquark with a scalar diquark and the ones with a vector diquark vary from $10-50$ MeV to $100-155$ MeV. 

Strange hidden-charm pentaquark states have rich spectroscopy, there are 10 negative-parity pentaquark states with $L=0$
and $25$ positive-parity pentaquark states with $L=1$ for each type of flavor configurations. The numerical results indicate that the masses of negative-parity ground states are distributed between $4200$ MeV and $4590$ MeV, while all pentaquark states with orbital excitation have masses above $4600$ MeV. The mass splitting between the S-wave and P-wave pentaquark states is about $350-570$ MeV.

A lowest strange hidden-charm pentaquark state with $J^P={1\over 2}^-$ was predicted around $4213$ MeV (denoted as $\Lambda_{c\bar c}(4213)$) in Ref.~\cite{PhysRevLett.105.232001, PhysRevC.84.015202}. The lowest strange hidden-charm pentaquark state with $J^P={1\over 2}^-$ was predicted around $4286$ MeV (denoted as $P'_{cs8}$) with an inclusion of the instanton-induced interaction~\cite{prd97.034006}, which has mass around $4367$ MeV without the instanton-induced interaction. In our numerical results, the lowest strange hidden-charm $[cs][\bar{c}qq]$ pentaquark state $| 1; 0, 1/2; 1/2, 0 \rangle_{1/2}$ with $J^P={1\over 2}^-$ has mass around $4200$ MeV.

From interior quark configurations and predicted masses, $P_{c\bar c s}(4338)^{0}$ observed by LHCb experiment could be assigned as the $|0; 1, 1/2; 1/2, 0\rangle_{1/2}$ pentaquark state in quark configuration~$[cq][\bar{c}sq]$ with $J^{P}=1/2^{-}$. Other three configurations with $J^{P}=1/2^{-}$ can not be ruled out under uncertainties of numerical results. Dynamical analysis reveals that the charm and anti-charm quarks belong to separate quark clusters (inter-cluster separation) in this configuration, which has a narrow decay width into $J/\psi \Lambda$.

$P_{cs}(4459)^{0}$ could be identified as the $|1; 0, 1/2; 3/2, 0\rangle_{3/2}$ pentaquark state in $[sq][\bar{c}cq]$ configuration. In this assignment, $P_{cs}(4459)^{0}$ has quantum numbers $J^{P}=3/2^{-}$. The presence of a spin-1 vector diquark within $P_{cs}(4459)^{0}$ leads to a mass approximately 120 MeV higher than that of $P_{c\bar c s}(4338)^{0}$. In this configuration, the $c$ and $\bar{c}$ quarks coexist within the same triquark cluster (intra-cluster coexistence), which results in a relatively broader decay width $\sim 17.3$ MeV in comparison to $\sim 7.0$ MeV for $P_{c\bar c s}(4338)^{0}$.

In the calculation, parameters in potentials are chosen from our previous work. However, the parameters $\alpha$ and $\lambda$ are fixed through limited ground tetraquark candidates, which will bring some uncertainties on the parameters. As pointed in Ref.~\cite{prd111.014015}, the parameter uncertainties may bring $\sim 10$ MeV uncertainties to masses of S-wave tetraquarks and $\sim 30$ MeV uncertainties to masses of P-wave tetraquarks, similar uncertainties will be brought to the masses of pentaquark states. In fact,how to determine all the parameters accurately and to find the interaction features behind deserves more investigation. For observed pentaquark candidates, there exist many other explanations as mentioned in the introduction. Except for non-resonance interpretations, there are many unsolved problems on resonance interpretations. As well known, the mixing among normal baryons, molecular pentaquarks and diquark-triquark pentaquarks will complex the spectroscopy of hadrons. How to distinguish different interior structures of pentaquark states is always a big challenge in both theory and experiment. There may exist many pentaquark states with different masses and $J^P$ quantum numbers in S-wave and P-wave excitations, how to find these pentaquark states from definite channels is also an experimental challenge.

\section*{Acknowledgements}
This work is supported by National Natural Science Foundation of China under Grant No. 11975146.


\bibliography{sn-bibliography}

\end{document}